# Critical processing temperature for high performance protected silver thin film mirrors

**David M. Fryauf,**[a,b] **Andrew C. Phillips,**[c] **Nobuhiko P. Kobayashi**[a,b,*]

[a]Electrical and Computer Engineering Department, Baskin School of Engineering, University of California Santa Cruz, Santa Cruz, California, United States, 95064

[b]Nanostructured Energy Conversion Technology and Research (NECTAR), Baskin School of Engineering, University of California Santa Cruz, Santa Cruz, California, United States, 95064

[c]University of California Observatories, University of California Santa Cruz, Santa Cruz, California, United States, 95064

**Abstract.** Silver (Ag) mirrors for astronomical telescopes consist of multiple metallic and dielectric thin films. Furthermore, the topmost surface of such Ag mirrors needs to be covered by a protection coating. While the protection coating is often deposited at room temperature and the entire mirrors are also handled at room temperature, various thin film deposition techniques offer protection coatings with improved characteristics when carried out at elevated temperatures. Thus, in this work, high-performance Ag mirrors were designed and fabricated with a new benchmark. The resulting Ag mirrors were annealed (i.e., post-fabrication annealing) at various temperatures to investigate the viability of introducing thermal processes during and/or after fabrication in improving overall optical performance and durability of protected silver mirrors. In our experiments, Ag mirror samples were deposited by electron-beam evaporation and subsequently annealed at various temperatures in the range from 60 °C to 300 °C, and then the mirror samples underwent an environmental stress test at 80 °C and 80% humidity for 10 days. While all the mirror samples annealed below 200 °C showed negligible corrosion after undergoing the stress testing, those annealed below 160 °C presented spectral reflectivity comparable to or higher than that of as-deposited reference samples. In contrast, the mirror samples annealed above 200 °C exhibited significant degradation after the stress testing. The comprehensive analysis indicated that delamination and voids caused by the growth of Ag grains during the annealing are the primary mechanisms of the degradation.

*Nobuhiko P. Kobayashi, Email: nkobayas@ucsc.edu

## 1 Introduction

High-performance astronomical telescopes benefit from durable broadband mirrors based on silver (Ag). Advantages of Ag include higher reflectivity and lower emissivity in the thermal infrared spectrum when compared to aluminum mirrors currently in prevalent use. There have been some successful implementations of Ag mirrors; notably those used on the Gemini telescopes[1,2,3]. However, few ground-based telescopes choose to utilize Ag mirrors in actual observatory environments, which reveals the elusiveness of Ag mirrors. The deep blue and UV spectra are very important for many astronomical research programs; however, even the Ag mirrors used in the Gemini telescope suffer from compromised deep blue and UV reflectivity due to the presence of an optically absorbing NiCrN adhesion film placed on a Ag film[4]. In addition, bare Ag tarnishes easily in the presence of oxygen and especially sulfur in Earth's atmosphere, and it experiences rapid corrosion via salt formation with halides. Therefore, reflective surfaces of Ag films must be immediately covered by an optically transparent protection coating to avoid tarnish and maintain original reflectivity for years.

Both the reflective Ag film and the subsequent protection coating are routinely deposited at room temperature by various physical vapor deposition (PVD) techniques including sputtering, evaporation, and cathodic arc deposition[5,6,7]. However, extensive literature indicates that PVD techniques produce protection coatings with superior durability against corrosion when carried out above room temperature[8,9,10,11]. Apart from PVD, atomic layer deposition (ALD) offers protection coatings with performance often superior to those deposited by PVD techniques. For instance, in our previous work, the highly uniform, conformal, and virtually pinhole-free nature of $Al_2O_3$ protection coatings deposited by ALD demonstrated superior durability over comparable $Al_2O_3$ deposited by electron-beam evaporation[12,13,14]. In our other previous work, it was found that ALD-based $Al_2O_3$ protection coatings on Ag exhibited higher durability with higher ALD processing temperature[15]. However, before further work can be done to maximize durability of Ag mirrors covered with protection coatings by ALD done at an optimal elevated processing temperature, the effects of such elevated processing temperatures on properties of Ag films themselves and of the entire Ag mirror structure must be thoroughly studied.

In this study, we report the effects of low-temperature (in the range of 60 °C to 300 °C) annealing on high-performance protected Ag mirrors designed with an innovative stack of thin films developed by a decade of previous work at the University of California Advanced Coatings Laboratory[16,17,18,19,20]. Bare Ag films and Ag mirrors covered with ultra-thin $Al_2O_3$ protective coatings were annealed at temperatures up to 500 °C and characterized to study the morphological changes. Although design specifications of most astronomical telescopes, including those of the Thirty Meter Telescope, discourage or prohibit any fabrication processes or maintenance procedures in which the mirrors are subjected to temperatures above 60 °C[21], our motivation for this present work is to understand the effects of low-temperature annealing on reflectivity and durability of protected Ag mirrors.

**2 Sample Preparation and Experimental Details**

Three types of Ag mirror samples, as illustrated in Figure 1, were deposited on 2.54 x 7.61 cm glass substrates using a custom e-beam evaporation tool, all of which share a common bottom adhesion layer of 22 nm $Y_2O_3$ on the substrate followed by a reflective film of 120 nm Ag. The first type shown in Fig. 1(a), referred to as "*UCO Ag 2018*", has an additional top adhesion layer of 10 nm NiO between the $Y_2O_3$ and Ag. *UCO Ag 2018* was covered with a three-layered protection coating: 2 nm NiO, 30 nm $YbF_3$, and 16 nm $TiO_2$. The second type illustrated in Fig. 1(b), referred to as "*Bare-Ag*" contains only a Ag film on the $Y_2O_3$ adhesion layer, while the third type shown in Fig. 1(c), referred to as "*$Al_2O_3$/Ag*", has a single film of 5 nm $Al_2O_3$ as a protection coating. The 5 nm thickness of the $Al_2O_3$ protection coating has been determined in our previous work as a bare-minimum film thickness which provides essential encapsulation protection to Ag films under ambient conditions. Each of the three types of mirror samples were deposited in vacuum with base pressure ~$10^{-6}$ Torr with no vacuum break. Ag films were deposited at a deposition rate of 4 nm/s while other films were deposited at 0.5 nm/s or slower. All listed thicknesses were obtained from quartz crystal microbalance.

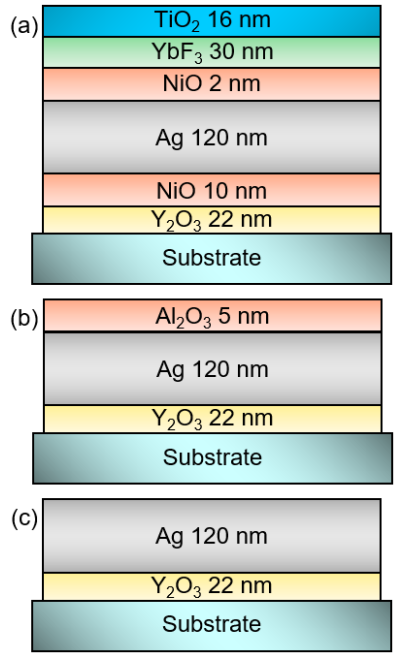

**Fig. 1** Illustration of the three types of Ag mirror samples showing details of layered structures with nominal thicknesses. (a) *UCO Ag 2018*, (b) *Bare-Ag*, and (c) *Al₂O₃/Ag*.

A set of fourteen identical coupons of *UCO Ag 2018*, including a control sample, was prepared and annealed in air for 30 minutes at anneal temperatures $T_a$ ranging from 60 °C to 300 °C with 20 °C intervals using a tube furnace (the control sample was not annealed). In addition, a set of *Bare-Ag* and a set of *Al₂O₃/Ag*, including control samples, were prepared; each set contained 6 identical coupons. Both *Bare-Ag* and *Al₂O₃/Ag* sets were annealed in air for 30 minutes at 100 °C intervals from 100 °C to 500 °C (the two control samples were not annealed). Specular reflectivity of all resulting mirror samples was measured at a 7-degree incident angle in the spectral range between 200 nm and 3 µm using a Cary 5000 spectrophotometer before and after annealing. Reflectivity spectra were quantified to a numerical value $R_{int}$ by integrating and normalizing the measured reflectivity between 300 nm and 1200 nm.

*UCO Ag 2018* coupons were subjected to environmental testing for ~10 days (231 hours) in a high-temperature high-humidity (HTHH) accelerated weathering environment described in

Phillips et al[22]. Briefly, the prepared coupons were arranged symmetrically and vertically in a desiccator jar with one edge mounted in custom Delrin® hardware. Wet KCl salt was loaded into the bottom of the jar to maintain relative humidity at ~75-80%, and the jar was loaded in an oven at 80 °C. The aggressive nature of the HTHH testing has been designed to accelerate various failure mechanisms that cause detectable failures within reasonable time so that at least some degree of degradation on even the best of coatings enables qualitative comparison between these specific mirror samples. After the HTHH testing, specular reflectivity of the mirror samples which retained specular reflecting area larger than the ~0.5 cm$^2$ spot size of the Cary 5000 spectrophotometer were re-measured. Durability of *UCO Ag 2018* samples that underwent the HTHH testing was quantified by evaluating the amount of remaining specular mirror area. The mirror samples were photographed with dark-field imaging by being illuminated from one side at an oblique angle by a halogen lamp to highlight the diffuse scattering from damaged features in contrast to the black background reflected by undamaged mirror areas. The collected dark-field images were then individually analyzed using the color saturation technique in ImageJ[23,21] software to calculate ratios of area that remained undamaged (i.e., area primarily showing specular reflection). *Bare-Ag* and *Al$_2$O$_3$/Ag* samples were further characterized by atomic force microscopy (AFM) and scanning electron microscopy (SEM) to evaluate surface roughness and morphology expected to change upon annealing. AFM scans were performed with a Bruker system in non-contact tapping mode under ambient conditions over a 5 um$^2$ area with frequency of 1 Hz. SEM images were taken with a FEI NOVA system at 50,000x magnification with beam settings of 10 kV and 0.13 nA.

**3 Results and Discussion**

The dark-field photographs of the *UCO Ag 2018* coupons that underwent the HTHH test are shown in Figure 2. Dark areas seen as a matte black background result from specular reflection of light, indicating the presence of areas undamaged during the HTHH testing, while light/colored areas result from scattering of light due to the presence of structurally damaged areas. Figure 2 shows a clear indication of an onset of damages that depends on $T_a$. The samples annealed up to 160 °C exhibit virtually no observable damages, while the samples annealed at 200 °C and above show that damaged areas increase as $T_a$ increases. The two samples annealed at 280 °C and 300 °C do not have a minimum undamaged area of ~0.5 cm$^2$ that allows meaningful reflectivity measurements.

$R_{int}$ of the HTHH-tested *UCO Ag 2018* samples annealed at various $T_a$ is plotted on the left axis in Figure 3. While $R_{int}$ appears not to be significantly impacted by varying $T_a$, the increased

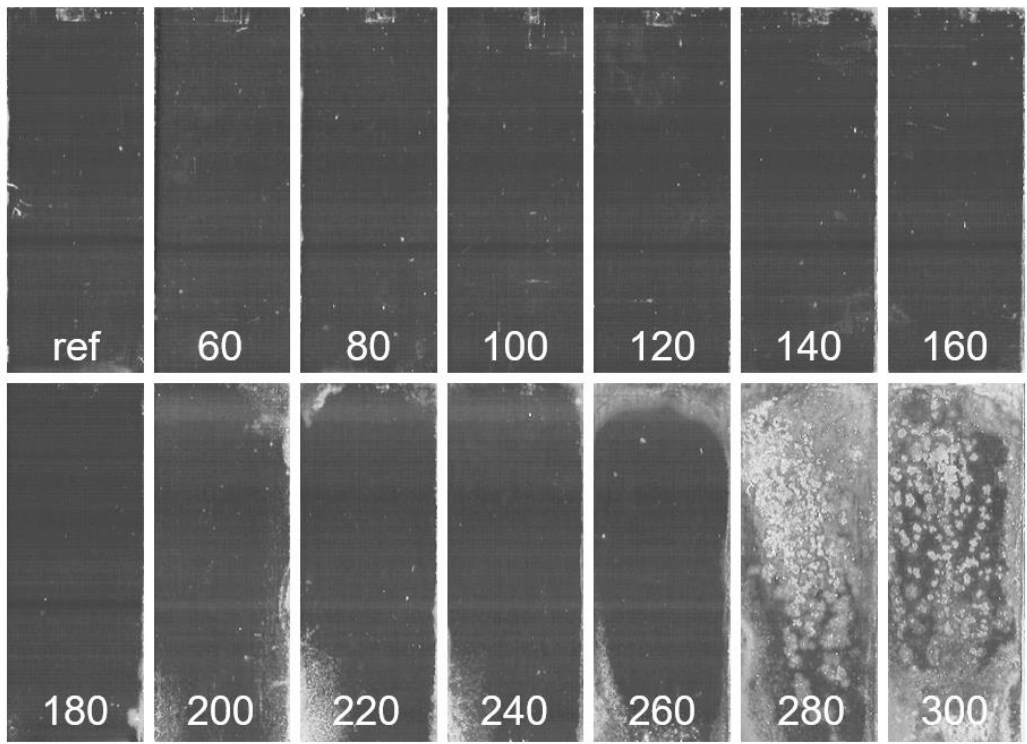

**Fig. 2** Dark-field photographs of the HTHH-tested *UCO Ag 2018* samples labeled in °C of $T_a$. Dark area represents undamaged specular mirror surface, and light/colored area represents diffuse scattering surface due to the presence of damages.

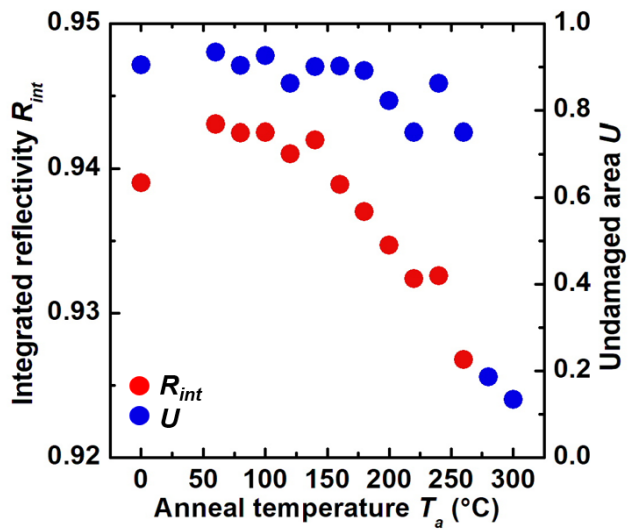

**Fig. 3** $R_{int}$ (red squares) and remaining undamaged specular reflecting area, $U$, (black circles) of the *UCO Ag 2018* samples.

$R_{int}$ of the samples annealed at temperatures in the range of 60 °C to 140 °C, relative to $R_{int}$ of the reference sample (i.e., the data point at $T_a$ of 25 °C), suggests that the mirror structure of *UCO Ag 2018* may benefit from post-fabrication annealing at temperatures in the range of 60 °C to 140 °C. The HTHH-tested *UCO Ag 2018* samples annealed at temperatures in the range of 160 °C to 260 °C show a clear trend of decreasing $R_{int}$ with increasing $T_a$, and $R_{int}$ of the HTHH-tested *UCO Ag 2018* samples annealed at 280 °C and 300 °C could not be measured due to the lack of the minimum undamaged area. Durability of the HTHH-tested *UCO Ag 2018* samples was further quantified by digital color saturation threshold analysis of the dark-field photographs in Figure 2. Remaining undamaged area of each sample was expressed as $U$ – the ratio of undamaged area to the total sample area – and is plotted on the right axis in Figure 3. The dependence of $U$ on $T_a$ shows behavior similar to that of $R_{int}$. In comparison to $U$ of the control sample (i.e., the data point at $T_a$ of 25 °C), the *UCO Ag 2018* samples annealed from 60 °C to 180 °C show no significant differences in $U$, and $U$ decreases with increasing $T_a$ from 200 °C to 300 °C. The similar trend seen in both $R_{int}$ and $U$ for the HTHH-tested *UCO Ag 2018* samples indicates that performance and durability of the *UCO Ag 2018* samples are at least not affected adversely by the annealing, and possibly improved by low-temperature annealing below 160 °C. The failure mechanisms – thermally-driven failure mechanisms – responsible for significant reduction in durability and $R_{int}$ of the HTHH-tested *UCO Ag 2018* samples that underwent post-fabrication annealing at temperatures above approximately 160 °C, as seen in Figure 3, were hypothesized to originate in the Ag thin film, which is expected to be most susceptible to undesirable structural rearrangements driven by thermal energy coupled with the post-fabrication annealing. To validate our hypothesis, two unique samples, *Bare-Ag* (Figure 1b) and *Al₂O₃/Ag* (Figure 1c), were prepared and analyzed to assess the dependence of structural changes that occur in the Ag thin film and related interfaces – parts in the *UCO Ag 2018* structure (Figure 1a) expected to be highly vulnerable to the post-fabrication annealing.

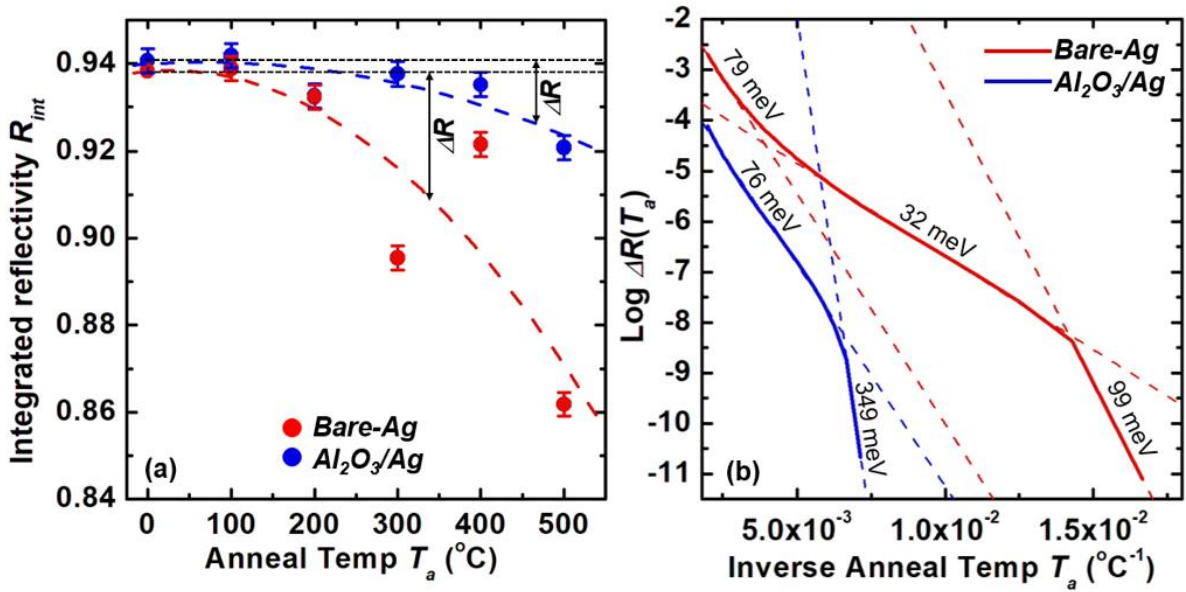

**Fig. 4** (a) $R_{int}$ of *Bare-Ag* and *Al₂O₃/Ag* samples after annealing at various $T_a$. Error bars represent the range in $R_{int}$ measurements of the as-deposited mirror sample set. (b) Log $R_{int}$ plotted as a function of the inverse of $T_a$. Activation energy $E_a$ extracted for various ranges of $T_a$ are denoted along the two lines – red and blue – that represent *Bare-Ag* and *Al₂O₃/Ag* samples, respectively.

Figure 4(a) shows $R_{int}$ of *Bare-Ag* and *Al$_2$O$_3$/Ag* samples annealed at 100 °C intervals up to 500 °C. Both sample sets annealed up to 500 °C exhibit a similar trend in $R_{int}$ at the low temperature range of interest up to 200 °C. However, the sharp decrease in $R_{int}$ of *Bare-Ag* above 200 °C, in contrast to *Al$_2$O$_3$/Ag*, indicates that the 5 nm Al$_2$O$_3$ protection coating substantially suppresses thermally-driven failure mechanisms of Ag thin films. A decrease in $R$ in the spectral range of 340~450 nm has been found to be the major contribution to decreased integrated $R$ for protected Ag mirrors[24,25,26,27]. This decreased $R_{int}$ in the near-UV/blue spectral range can be attributed to losses at the top reflecting Ag film interface, due to increased light scattering caused by morphological roughness and increased light absorption associated with surface plasmon polariton resonance at roughened Ag/protection coating interfaces. In Figure 4(a), the red and blue dotted lines approximate the two different trends seen in *Bare-Ag* and *Al$_2$O$_3$/Ag* sample sets (Note: the two data points at $T_a$ = 0 °C represent two reference samples that were not subjected to annealing).The two trends seen in Figure 4(a) are interpreted as two functional forms of $T_a$ to extract a difference *DR* between $R_{int}$ of each of the two samples (i.e., *Bare-Ag* and *Al$_2$O$_3$/Ag*) and that of their respective reference samples at various $T_a$. In Figure 4(b), natural logarithm of resulting *DR* is plotted as a function of the inverse of $T_a$ to extract activation energies $E_a$ – presumably associated with various degradation mechanisms manifested as decreases in $R_{int}$ – that emerge in various ranges of $T_a$. Three noticeable features arise in comparing the two samples. First, *Bare-Ag* exhibits three discernible ranges – low, middle, and high denoted with $E_a$ of 99 meV, 32 meV, and 79 meV, respectively – of $T_a$. In contrast, *Al$_2$O$_3$/Ag* samples appear to unveil only two distinguishable ranges – low and high denoted with $E_a$ of 349 meV and 76 meV, respectively – of $T_a$. Although detailed structural and chemical assessment is necessary to explicitly illustrate specific degradation mechanisms manifested in a distinctive range of $T_a$, qualitative and comparative assessment based on Figure 4(b) is ventured. In the high $T_a$ range, $E_a$ of the two samples, 79 meV for *Bare-Ag* and 76 meV for *Al$_2$O$_3$/Ag*, is comparable; thus, the degradation mechanism is likely to be directly associated with deteriorations that take place within the Ag films. In the low $T_a$ range, $E_a$ of the two samples, 99 meV for *Bare-Ag* and 349 meV for *Al$_2$O$_3$/Ag*, is substantially different. In addition, the two onsets, $1/T_a \sim 1.7 \times 10^{-3}$ °C$^{-1}$ for *Bare-Ag* and $7.5 \times 10^{-2}$ °C$^{-1}$ for *Al$_2$O$_3$/Ag*, obtained by extrapolating their respective line segments onto the x-axis are significantly different from one another. These marked differences seen in the low $T_a$ range qualitatively suggest that the *Al$_2$O$_3$/Ag* structure be considerably less vulnerable to the inception of unique degradation mechanisms. More importantly, the *Al$_2$O$_3$/Ag* structure is immune up to $T_a$ much higher than that at which detectable degradations commence in the *Bare-Ag* structure. The presence of the middle $T_a$ range for *Bare-Ag* may indicate that such transitional degradation mechanisms as those occurring on the surface of unprotected Ag thin films play a role.

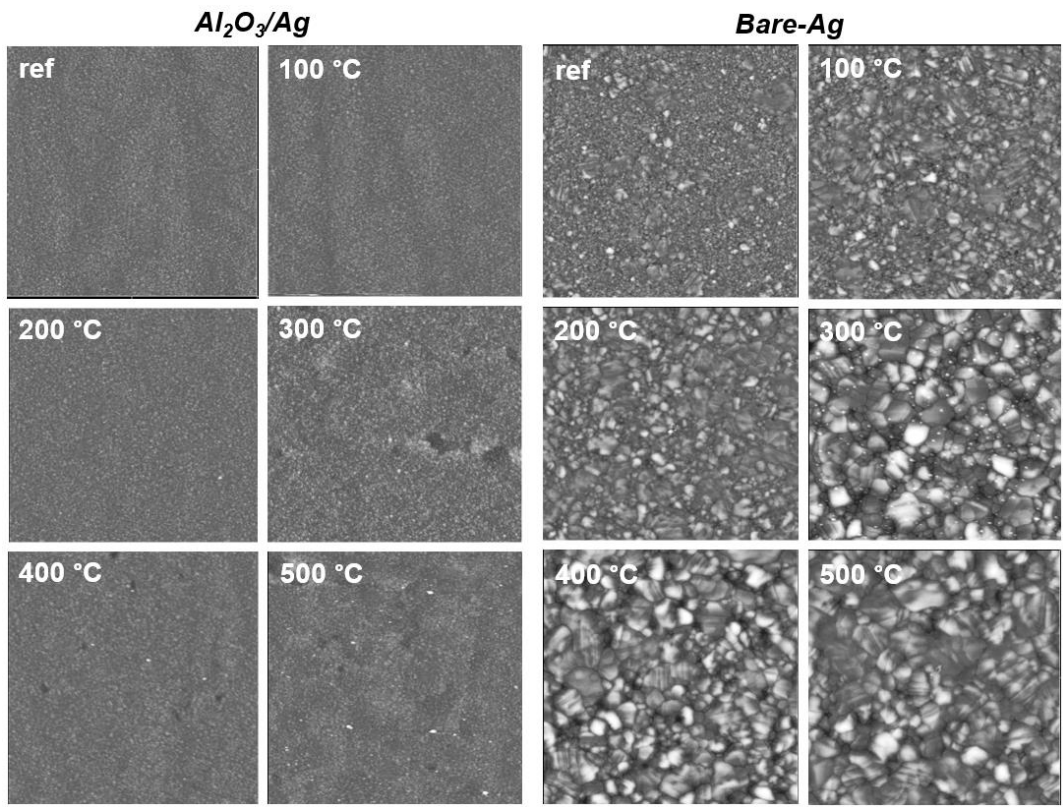

**Fig. 5** 5 µm$^2$ AFM scans of *Al$_2$O$_3$/Ag* and *Bare-Ag* sample sets after annealing. The z-axis range of all scans is normalized to 100 nm.

AFM scans of the annealed *Bare-Ag* and *Al₂O₃/Ag* samples were acquired, shown in Figure 5, to observe morphological roughening of Ag surfaces as $T_a$ was increased. The AFM scans, normalized to a common Z-axis range of 100 nm, portray an obvious morphological evolution of surfaces for the *Bare-Ag* samples. As $T_a$ increases so does surface feature size and overall magnitude of surface roughness. In contrast, the *Al₂O₃/Ag* samples show morphological evolution much less evident than those seen on the *Bare-Ag* samples, although some ambiguous features including pits emerged on the surface of the *Al₂O₃/Ag* sample annealed at 300 °C. Surface roughness displayed in Figure 5 is quantified and plotted in Figure 6 to better illustrate the contrast in the dependence of surface roughness on $T_a$ between the two sample sets. The trend seen in Figures 5 and 6 clearly suggests that the 5 nm thick $Al_2O_3$ protection coating significantly suppresses surface roughening on Ag surfaces during post-fabrication annealing. AFM can only

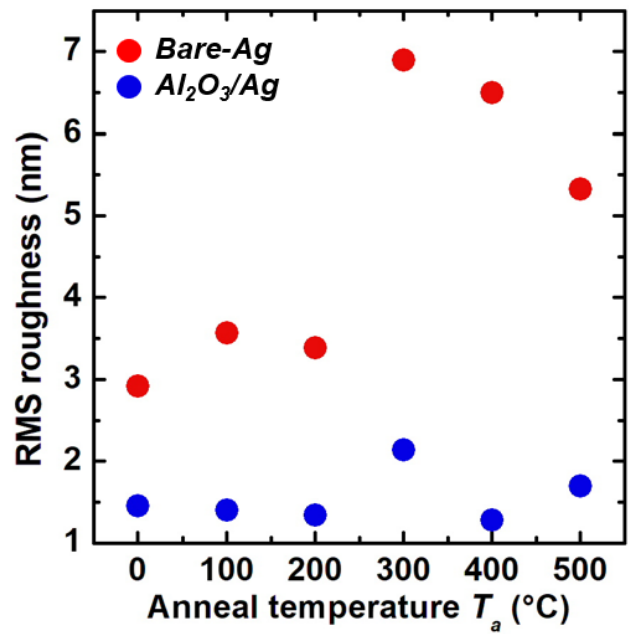

**Fig. 6** RMS roughness from 5 µm² are of *Bare-Ag* and *Al₂O₃/Ag* sample sets after annealing at various $T_a$.

characterize a topmost surface of a sample; it cannot directly observe the Al$_2$O$_3$/Ag interface. In other words, Figures 5 and 6 confirm that the Al$_2$O$_3$ protection coating suppresses the formation of features that develop vertically (i.e., along the z-direction) – features measured as surface roughness – however, AFM provides almost no insight into probable lateral diffusion of Ag and grain growth in the Ag film. Therefore, the SEM images shown in Figure 7 are used to complement and supplement the AFM characterization from Figures 5 and 6. The accelerated electron with energy of 10 keV was used in SEM imaging so that electrons adequately tunnel through the 5 nm Al$_2$O$_3$ protection coating, allowing the electrons to interact with the underlying Ag interface with minimal scattering[28,29]. Although plan-view SEM imaging, unlike AFM imaging, cannot be used to quantify morphological information along the z-axis, SEM often offer improved lateral spatial resolution and boosted imaging contrast of surfaces – enhanced SEM imaging – with high electrical conductivity, such as Ag films, when compared to that of AFM imaging often skewed by the tip-surface convolution[30].The enhanced SEM imaging was used to image the as-deposited

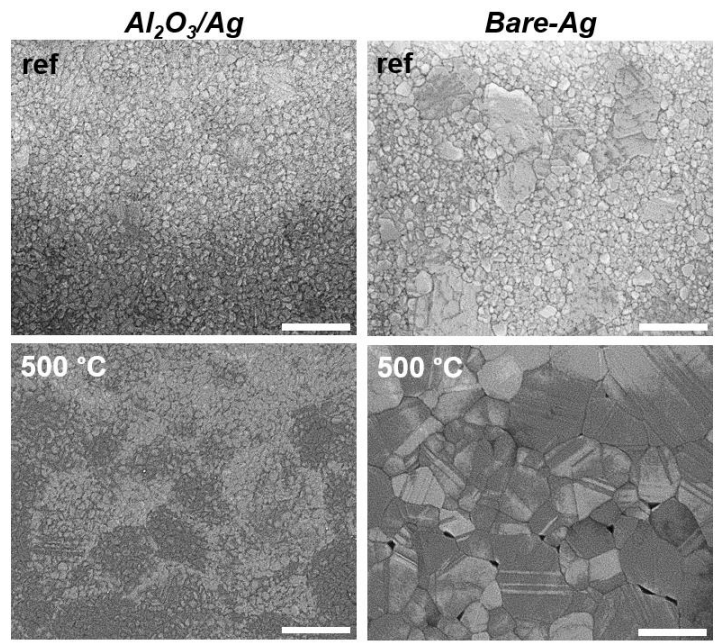

**Fig. 7** SEM images of as-deposited and 500 °C-annealed *Bare-Ag* and *Al$_2$O$_3$/Ag* samples. Scale bar represents 500 nm.

*Bare-Ag* and *Al₂O₃/Ag* samples (i.e., samples with no annealing) and the 500 °C-annealed *Bare-Ag* and *Al₂O₃/Ag* samples as shown in Figure 7, displaying the vivid evolution of the Ag films after annealing. The as-deposited samples, Figures 7(a) and 7(b), exhibit grain size distribution dominated by smaller (~50 nm diameter) features with limited presence of larger (~200-500 nm diameter) features. In contrast, both *Bare-Ag* and *Al₂O₃/Ag* samples, Figures 7(c) and 7(d), annealed at 500 °C exhibit grain size distribution predominantly composed of larger (~200-500 nm diameter) features. Occasional voids are observed in the *Bare-Ag* sample annealed at 500 °C, represented by dark spots in Figure 7(c), which can be attributed to the de-densification of grain boundaries as thermally-driven diffusion during the annealing is expected to cause Ag grains to grow both vertically and laterally. In contrast, the *Al₂O₃/Ag* sample annealed at 500 °C, Figure 7(d), displays Ag grains much larger, on average, than those seen in Figure 7(c), suggesting that lateral growth has been promoted while vertical growth has been suppressed or even negated as the total amount of Ag was conserved. The grain boundaries under the 5 nm $Al_2O_3$ protection coating in Figure 7(d) appear to have brightness higher than that of the interior of the grains while the grain boundaries seen in Figure 7(c) have brightness lower than that of the interior of the grains, indicating that the grain boundaries in the panel (d) would impede transportation of electrons from one grain to another and would also suppress mass transport of Ag from one grain to another. Furthermore, the grain boundaries with brightness higher (i.e., electrical conductivity lower) than that of the interior of grains, in Figure 7(d), may indicate the presence of delamination and/or voids under the $Al_2O_3$ protection coating. Several comparable studies on the micro-structure and morphology of protected Ag mirrors experiencing corrosion have identified delamination and voids between Ag films and protection coatings as a common indication of failures and resulting structural damages[31,32,33,34].

Both *Bare-Ag* and *Al₂O₃/Ag* samples are expected to experience increased Ag grain size, decreased grain boundary density, and film densification during annealing. The comparison of surface roughness between *Bare-Ag* and *Al₂O₃/Ag* samples annealed at 500 °C in Figures 5 and 6 suggests that the 5 nm Al₂O₃ protection coating suppresses surface roughening during annealing. Nevertheless, the comparable increase in grain size of the two samples, *Bare-Ag* and *Al₂O₃/Ag*, in Figure 7 suggests that the fundamental mechanism of increased surface roughness resulting from diffusion-driven grain growth occurs within the Ag films independent of the presence of a Al₂O₃ protection coating. The series of processes – temperature-induced grain growth, grain boundary de-densification, film densification and resulting surface roughening – is a well-understood and expected effect seen in thin metallic films that undergo annealing[35,36,37]. Grain growth during annealing proceeds by rearranging existing atoms (i.e., mass is conserved). If grain growth is allowed to continue in three-dimensions (i.e., in a Ag film with no protection coating), it proceeds in all three directions: the in-plane and the z-direction. In contrast, if grain growth is limited to the in-plane by a protection coating, it proceeds only laterally, making more Ag atoms available for the grain growth. The characterization presented in Figures 5 – 7 suggests that grain growth is thermally driven in the Ag film with a single-layer Al₂O₃ protection coating; thus, even with more complex designs of protection coating (e.g., the one used in the *UCO Ag 2018* samples), Ag films are expected to experience such grain growth during annealing conducted above a certain critical temperature. Furthermore, we suggest that delamination and voids, such as those observed in Figures 5(d)-(f) and Figure 7(d), most likely to be induced by grain growth in the Ag films, are responsible for both the optical and mechanical degradation that occurred during the HTHH tests. The range of $T_a$ beneficial to optical Ag thin films should be more thoroughly assessed in further experiments. As-deposited Ag films exhibit smooth surfaces made of small average grain size as shown in Figures 5(a) and 5(b). However, our series of experiments clearly indicated that the two specific mirror structures, *UCO Ag 2018* and *Al$_2$O$_3$/Ag*, sustained negligible adverse effects as long as they were annealed at temperature below 160 °C. Therefore, we propose that 160 °C be used as the highest temperature that can be used in post-fabrication-annealing for Ag thin films covered with protection coatings with the goal of improving both *R* and durability.

**4 Conclusion**

A new protected Ag mirror design, *UCO Ag 2018*, was introduced and used to assess risks and benefits of post-fabrication annealing in terms of its $R_{int}$ and environmental durability. No detrimental effects were observed for annealing carried out at temperatures below 160 °C. The assessment was further supplemented by studying simpler structures: *Bare-Ag* and *Al$_2$O$_3$/Ag*. Failure mechanisms, which consistently decreased $R_{int}$ as $T_a$ was raised from 160 °C to 260 °C and eventually caused catastrophic failure at temperatures above 260 °C, were analyzed with the *Bare-Ag* and *Al$_2$O$_3$/Ag* samples. Formation of void defects and delamination between the Ag thin film and adjacent barrier film, caused by Ag grain growth during annealing at temperatures higher than 160 °C, are suggested as the major failure mechanisms. Our experiment suggests that the common practice, for Ag-based protected mirrors, of not using post-fabrication processing at elevated temperature should be re-evaluated for the sake of improving performance and durability.

*Acknowledgements*

We thank Brian Dupraw for depositing the mirror samples at the University of California Observatories Advanced Coatings Lab. We also thank Hewlett-Packard Laboratories (Palo Alto, California) for letting the authors use AFM and SEM. We also thank the following two NSF grants, 1407353 (Dr. Zoran Ninkov) and 1562634 (Dr. Thomas F. Kuech), for partial financial support.